\documentclass[12pt]{article}
\usepackage{amssymb,amsmath}
\begin{document}

\renewcommand{\thefootnote}{\fnsymbol{footnote}}

\begin{center}
{\large {\bf Spin-Isospin Rotation Dynamics
 \\[0pt] }}
 \end{center}

\begin{center}
\textbf{V. D. Tsukanov}
\end{center}
\begin{center}{\it Institute of Theoretical Physics, National Science Center}\\[0pt]
{\it ''Kharkov Institute of Physics and Technology''}\\[0pt]
{\it 61108, Kharkov, Ukraine}\\[0pt]
\end{center}

\begin{quotation}
{\small {\rm The equations for the solitons arbitrarily rotating
in the ordinary and isotopic space are obtained. The wave
functions of the corresponding dynamic states in the quantum case
are found. The generalized matrix of the moments of inertia is
degenerate for the O(2)-invariant configurations characteristic
for the nucleon and delta-isobar. The equation for such
configurations is established. It is shown that the spin-isospin
rotation prevents the collapse of the soliton states in the SU(2)
sigma-model. The entire consideration is based on the variational
approach to the method of collective variables.}}

\end{quotation}

\renewcommand{\thefootnote}{\arabic{footnote}} \setcounter{footnote}0
\vspace{1cm}

\section{Introduction} \label 1

Ordinary and isotopic spins are the fundamental characteristics of nucleons.
Therefore the key moment of any realistic theory of nucleons is the search for
suitable localized states performing rotations in the ordinary and isotopic
spaces. The existing papers describe such states with the help of the
adiabatic approximation. Such an approach uses static soliton solutions as a
trial ansatz whereas the degeneration parameters of these solutions play the
part of collective coordinates. This procedure, also known as the moduli space
approximation, found wide application in the field theory. Specifically, in
the Skyrme model it was employed in quantizing the hedgehog solution
\cite{ANW}, as well as in quantizing the soliton states in higher
homotopic classes \cite{Ir}.

At the same time, there exist actual problems whose solution within the
framework of the adiabatic approach is impossible. It concerns the exact
account of the dynamic deformation of a nucleon, the description of the
spectrum of diverse rotational resonances, and the search for the localized
states in the SU(2) sigma-model. The formulation of the problems mentioned
requires the extended set of collective variables. Their choice, first of all,
must reflect the symmetry properties of the initial Lagrangian. This will
permit to consider the solitons rotating independently in the ordinary and
isotopic spaces. The configuration of the localized states with such an
approach must from the very outset be determined with the account for their
collective dynamics characterized by the certain values of the ordinary and
isotopic spins. These requirements do not assume the obligatory existence of
the static soliton solutions. Therefore their realization can revitalize the
SU(2) sigma-model as a potential model of nucleons. It is known that in a
topological sector due to the Derrick theorem the static soliton solutions
degenerate into phantom point states with zero energy. If such states are
rotated, they acquire finite size and finite energy. As regards the stability
of rotating states, this problem arises when considering any models, and the
sigma-model does not differ in this respect from any other. If the
quantization scheme with a half-integer spin is used, just a rotation will
correspond to the ground state of such a soliton. In order to stabilize the
moderate rotations with respect to a disintegration, a mass term should be
added to the Lagrangian to provide for the damping asymptotics of the rotating
states at the infinity \cite{BR}. A consistent study of dynamic soliton
states can be performed within the framework of the collective variables
formalism.

The exact equations of the collective dynamics are a natural consequence of
the Hamiltonian equations of the total system. Let $X, P$  be a set of
canonical collective variables reflecting the characteristic properties of
the problem considered, and $q, p$ be all other variables tentatively called
microscopic. The exact temporal evolution of the total system is described
with the help of the Hamiltonian equations
\begin{equation}
\begin{split}
&\frac{\partial H(X,P;q,p)}{\partial p} =\dot{q},\;\;\;\,\frac{%
\partial H(X,P;q,p)}{\partial q}=-\dot {p}, \\
&\frac{\partial H(X,P;q,p)}{\partial P} =\dot{X},\;\;\;\,\frac{%
\partial H(X,P;q,p)}{\partial Q}=-\dot{P}.  \label{p}
\end{split}
\end{equation}
The assumption that the influence of microscopic variables on the collective
subsystem is effective only on the time scale considerably exceeding the
characteristic time of the own evolution of the collective subsystem forms the
ground for using the method of collective variables. In other words, the
collective subsystem is actually conservative within a sufficiently broad time
interval. In order to separate the principal, governing motions and to exclude
completely the minor influence of microscopic variables under these conditions,
they should be frozen by putting $\dot{q}=0,\, \dot{p}=0$. This requirement
transforms the complete set of the Hamiltonian equations (\ref p) into the
exact equations of the collective dynamics
\begin{equation}
\frac{\partial H(X,P;q,p)}{\partial p}=0\,,\;\;\;\,\frac{\partial H(X,P;q,p)}
{\partial q}=0\,,  \label{g}
\end{equation}
\begin{equation*}
\frac{\partial H_{c}(X,P)}{\partial P}=\dot{X},\;\;\;\,\frac{\partial H_{c}
(X,P)}{\partial Q}=-\dot{P}.
\end{equation*}
The first pair of these equations determines the coherent components of the
microscopic variables $\overline{q}\equiv \overline{q}(X,P)$, \ $\overline{p}
\equiv \overline{p}(X,P)$, that minimize the complete Hamiltonian of the
system with fixed values of the collective variables $X,P$. The second pair
represents the Hamiltonian equations for the collective subsystem. The
collective Hamiltonian $H_{c}(X,P)$ is defined as the complete Hamiltonian of
the system dependent on the coherent components of the microscopic variables:
\begin{equation*}
H_{c}(X,P)\equiv H(X,P;\overline{q}\,(X,P),\overline{p}\,(X,P))\,.
\end{equation*}
Obviously, in such an approach the complete Hamiltonian of the
system is presented in the form of the expansion in powers of the
fluctuations of microscopic variables
$\widetilde{q}=q-\overline{q}\,(X,P)$, \ \ $\widetilde{p}=
p-\overline{p}\,(X,P)$. The zero term of such an expansion
coincides with the collective Hamiltonian, and the account of the
fluctuations starts with the quadratic terms. Thus, the approach
given provides for the complete description of the total system in
the microscopic sense reflecting at the same time the priority of
the collective subsystem. The scheme presented clearly
demonstrates the variational nature of the method of collective
variables noted in the paper \cite{Ts1}.

The goal of this paper is to study the dynamic solitons in SU(2)
field theories. The general questions of the collective
description such as gauging, changing of variables, and
variational equations are outlined in Sec. \ref2 by way of example
of the systems described by the Lagrangian quadratic in velocity.
Sec. \ref3 deals with the field SU(2)-theories. The equations for
the solitons performing arbitrary rotations in the ordinary and
isotopic space are obtained. Sec. \ref4 discusses the link between
the equations of collective dynamics and the exact self-similar
solution of the equations of motion. Sec. \ref5 presents the wave
functions of arbitrary dynamic states obtained under spontaneous
breakdown of symmetry entangling spin and isotopic variables. Sec.
\ref6 performs the transition from the general equations for the
states with arbitrary spin and isospin values to the equations for
the configurations with the axial symmetry. Such configurations
describe the nucleon and delta-isobar states. The transition noted
is not obvious because the matrix of the generalized moments of
inertia is degenerate in case of axial symmetry. It is shown in
this section that accounting for rotations in the SU(2)
sigma-model impedes the collapse of the soliton states
characteristic for the static solutions. In Conclusion the main
results obtained in this paper are briefly summarized. From the
viewpoint of interpretation and physical sense the procedure of
the canonical description used in this paper differs from the
similar procedure applied within the framework of the adiabatic
approximation \cite{ANW}. However, from the technical viewpoint,
the elements of this description are practically identical. These
questions are outlined in the Appendix.

\section{Equations of collective dynamics} \label 2

The transition to the collective description is accompanied by the change of
variables in the configurational space. The choice of microscopic variables
for such a change is not unique. Using this arbitrariness in a proper way can
considerably simplify the equations of the variational approach as well as the
equations of motion for the fluctuations. In order to demonstrate these and
other elements of the variational approach that are common to a broad class of
models, let us consider the nonlinear system described by a nondegenerate
Lagrangian quadratic in velocity
\begin{equation*}
L=\frac{1}{2}\sum g_{ik}(q)q_{,\,t}^{i}q_{,\,t}^{k}-H(q,0)\,.
\end{equation*}
Here $q^i$ are the generalized coordinates of the system. The
approach outlined below is equally applicable to dynamic systems
with a finite number of degrees of freedom as well as to the field
theory. As applicable to the field systems, $i\equiv \{i,\ x\}$ is
the set of discrete indices and spatial coordinates. It is
expedient to present the transition to new variables as follows.
We will regard as new microscopic variables the initial
coordinates $q^i$, that simultaneously become the functions of the
limited set of collective variables $Q^{\alpha }$:
$q^{i}=q^{i}(Q)$ . In order to provide for the nondegenerate
nature of the suggested change of variables, we limit the
admissible variations of the microscopic variables $\delta
q^{i}(Q=const)$ by the orthogonality conditions
\begin{equation}
\frac{\partial q_{0}^{i}}{\partial Q^{\alpha }}\,g_{ik}(q_{0})\,\delta
q^{k}=0\,. \label{c}
\end{equation}
Here $q_{0}^{i}\equiv q_{0}^{i}(Q)$ is the trial ansatz depending
only on collective variables whose form will be defined later.
Obviously, the configurational space of the system is considered
as a Riemann manifold with the metrics $g_{ik}(q)$ generated by
the kinetic term. Let us introduce the projection operator on the
subspace formed by the tangent vectors $\partial
q_{0}^{i}/\partial Q^ {\alpha}$:
\begin{equation}
\mathcal{P}_{k}^{i}=g_{ks}(q_{0})\frac{\partial
q_{0}^{s}}{\partial Q^{\alpha }}g^{\alpha \sigma
}(q_{0})\frac{\partial q_{0}^{i}}{\partial Q^{\sigma }}\,,\qquad
\mathcal{P}^{2}=\mathcal{P}\,,  \label{n}
\end{equation}
where $g^{\alpha \sigma }(q_{0})$ is the matrix inverse to the tensor
\begin{equation}
g_{\alpha \sigma }(q_{0})=\frac{\partial q_{0}^{i}}{\partial Q^{\alpha }}%
g_{ik}(q_{0})\frac{\partial q_{0}^{k}}{\partial Q^{\sigma }}\,,  \label{k}
\end{equation}
 defining the metrics on the surface of the submanifold $q_{0}^{i}(Q)$. In
terms of the operator $\mathcal{P}$ the constraint condition (\ref c) can be
presented in the form
\begin{equation}
\delta q(1-\mathcal{P})=\delta q\,.  \label{e}
\end{equation}
The variations in this relation can be ridden by a time dependence
of microscopic variables $\delta q=\dot{q}\delta t\, (Q=const)$.
Therefore the velocities of microscopic variables and their
variations will also satisfy to similar constraint conditions
\begin{equation}
\dot{q}(1-\mathcal{P})=\dot{q},\qquad \delta
\dot{q}(1-\mathcal{P})=\delta \dot{q}\,.  \label{f}
\end{equation}
 Noting that
\begin{equation*}
q_{,t}=\dot{q}(Q)+\dot{Q}\frac{\partial q(Q)}{\partial Q}
\end{equation*}
and accounting for Eq. (\ref f) in calculating the variational derivatives,
let us find the canonical momenta conjugated to new variables
\begin{eqnarray}
&P_{\alpha }&\equiv \frac{\partial L}{\partial \dot{Q}^{\alpha }}=\frac{%
\partial q^{s}}{\partial Q^{\alpha }}g_{sk}(q)\left( \frac{\partial
q^k}{\partial Q^\sigma}\dot{Q}^\sigma+\dot{q}^k\right)\,,  \label{j} \\
&p_{\,i}&\equiv \frac{\partial L}{\partial
\dot{q}^{i}}=(1-\mathcal{P})_i^s g_{sk}(q)\left( \frac{\partial
q^k}{\partial Q^\sigma}\dot{Q}^\sigma+\dot{q}^k\right)\,.
\label{h}
\end{eqnarray}
In these relations the coefficients in front of the velocities
$\quad \dot{Q}^ {\alpha }$, \ $\dot{q}^{k}$ define the elements of
the block kinetic matrix $G(q)$. Writing the Hamiltonian of the
system in new variables
\begin{equation}
H=\dot{q}\frac{\partial L}{\partial \dot{q}}+\dot{Q}\frac{\partial L}{%
\partial \dot{Q}}-L=\frac{1}{2}(P,\ p)\,G^{-1}(q)\left(
\begin{array}{c}
P \\
p
\end{array}
\right) +H(q,0)\,,  \label{m}
\end{equation}
we can turn to the equations for the coherent components
$\overline{p}$, $\overline{q}$ (\ref g). Accounting for the
constraint $p(1- \mathcal{P})=p$ (\ref h), the first of them
becomes
\begin{equation}
(0,(1-\mathcal{P}))\,G^{-1}(\overline{q})\left(
\begin{array}{c}
P \\
\overline{p}
\end{array}
\right) =0\,.  \label{i}
\end{equation}
In an extremum point, according to Eq. (\ref g), the arbitrary
microscopic coordinates $\varrho$ become the functions of the
collective variables $\overline{\varrho}(Q,P)$. Therefore, in an
extremum point the collective variables dependence of the initial
coordinates $q\equiv q(Q,\varrho)$ can be presented in the form
$\overline{q}(Q,P)=q(Q,\overline{\varrho}(Q,P))$. Generalizing the
approach offered in the paper \cite{Ts2}, let us define the gauge
function $q_0$ and its derivatives in (\ref c) as the limits
\begin{equation}
q_0(Q)=q(Q,\varrho)_{|\varrho\rightarrow\overline\varrho(Q,P)}=\overline{q}(Q,P),
\qquad \frac{\partial q_0(Q)}{\partial Q}={\frac{\partial
q(Q,\varrho)}{\partial
Q}}_{|\varrho\rightarrow\overline\varrho(Q,P)}.  \label{xy}
\end{equation}
Due to this definitions and the constraints (\ref c), (\ref f) the
nondiagonal blocks of the kinetic matrix $G(\overline q)$
(\ref{j}), (\ref{h}) vanish in an extremum point. So the matrix
itself splits into two independent blocks $g_{\alpha \sigma}(\
\overline{q})$ and
$(1-\mathcal{P})_{i}^{s}g_{sk}=g_{is}(1-\mathcal{P})_{k}^{s}$,
relating to the collective and microscopic subsystems,
respectively. Eq. (\ref i) comes homogeneous  with respect to
$\overline{p}\,$, and we obtain $\overline{p}=0$. It is the main
result of this section. So, in the used gauge the equation
determining the coherent components of the coordinates $q$ and the
corresponding functional $H_c(q,P)$ will have the forms
\begin{equation}
\frac{\partial H_{c}(q,P)}{\partial ^{\,\prime }q^{i}}=0,\qquad
H_{c}(q,P)\equiv \frac{1}{2}\,g^{\alpha \sigma }(q)P_{\alpha
}P_{\sigma}+H(q,0). \label{xx}
\end{equation}
Here, in contrast to the ordinary derivative $\partial /\partial
q$, the derivative over $q$ accounting for the constraint (\ref e)
is denoted with a symbol $\partial /\partial ^{\,\prime }q$. If
the collective coordinates are cyclic ones, the dependence of the
initial variables from these coordinates is determined obviously.
It allows easily to establish the structure of the equation (\ref
{xx}). Just such a case is considered further in this paper. In
general case, to find the equation for the coherent components
$\overline q$, let us consider the variation of the functional
$H_{c}(q,P)$ with respect to the variations of microscopic
variables $\delta q$ at $Q,P$ fixed. Since the Hamiltonian
$H_{c}(q,P)$ is the function of the variables $q$ and $q_{,\alpha
}\equiv \partial q/\partial Q^{\alpha }$ and accounting that under
this conditions $\delta(q_{,\alpha})=(\delta q)_{,\alpha}$, we
obtain
\begin{equation*}
\delta H_{c}(q,P)=\delta q^k \frac {\partial H_c}{\partial
q^k}+\frac {\partial}{\partial Q^{\alpha}}(\delta q^k)\frac
{\partial H_c}{\partial q_{,\alpha}^k}\,.
\end{equation*}
Carrying out all differentiations over $Q$ at
$\overline\varrho(Q,P)$ fixed, let us change variables $q\equiv
q(Q,\varrho)$ by their coherent components
$q(Q,\overline\varrho(Q,P))$. Then moving the derivative over $Q$
and noting that in an extremum point
\begin{equation*}
\delta q^k\frac {\partial H_c}{\partial q_{,\alpha}^k}=0\,,
\end{equation*}
we obtain the basic equation of the variational approach
\begin{equation}
(1-\mathcal{P})_{i}^{k}\left( \frac{\partial H_{c}}{\partial q^{k}}-\frac{%
\partial }{\partial Q^{\alpha }}\frac{\partial H_{c}}{\partial q_{,\alpha }^{k}}%
\right) =0\,.  \label{o}
\end{equation}
Let us stress that this equation determines the coherent component
of the coordinate, and therefore, in view of the choice of the
gauge function, one should change the ansatz $q_{0}$ in the
projection operator $\mathcal{P}$ for the sought function $q$
(\ref{xy}). In the field systems the equation (\ref o) describes
the configurations of dynamic solitons. Within limit of $P_{\alpha
}=0$ it reduces to the equation for the quasistatic configurations
\begin{equation*}
(1-\mathcal{P})_{i}^{k}\frac{\partial H(q,0)}{\partial q^{k}}=0\,.
\end{equation*}
This equation determines static stresses of deformed solitons and
in particular allows to establish properties of the intersoliton
potentials \cite{Ts3}. The Hamiltonian $H_{c}(q,P)$ dependent on
the components $\overline{q}(Q,P)$ defines the collective
Hamiltonian of the system:
\begin{equation*}
H_{c}(Q,P)\equiv \frac{1}{2}\,g^{\alpha \sigma}(\overline{q}(Q,P))
P_{\alpha }P_{\sigma }+H(\overline{q}(Q,P),0)\,.
\end{equation*}
Performing the further expansion of the Hamiltonian (\ref m) in powers of the
fluctuations of microscopic variables we can go beyond the framework of the
purely collective description. The inclusion of the higher-order terms permits
to determine the fluctuation spectrum as well as the effect of fluctuations on
the collective motion.

Note that the equations obtained are invariant with respect to the
general gauge transformation of collective variables $Q\rightarrow
Q^{\prime }(Q)$. Specifically, this is seen from the structure of
the projection operator $\mathcal{P}$ (\ref n), (\ref k) that is
invariant with respect to such transformations. As regards the
microscopic variables, their choice is fixed through the gauge
condition (\ref c). The invariance properties of the theory with
respect to the choice of microscopic parameters are not considered
in this paper. We only mention a concrete example of such
invariance. It is shown in the paper \cite{Ts2} that the zero mode
in the fluctuation spectrum of the one-dimensional problem are
removed automatically owing to the inherent properties of the
theory and regardless of the gauge form used. So this result
solves the "zero-mode problem" appeared in the middle of the
seventies in connection with the application of the perturbation
theory to the soliton systems.

The gauge (\ref c) used in this paper, in which the gauge function
is identified with the extremal of the Hamiltonian (\ref{xy}),
reduces to zero the coherent component of the canonical momentum
and factorizes the kinetic matrix at the extremum point. All this
simplifies the analysis of the equations for the coherent
components as well as equations for fluctuations. Using similar
gauges when considering concrete systems permits to formulate a
simplified receipt for describing a purely collective motion. That
is, using the Lagrange formalism, one can omit the velocities of
microscopic variables from the very beginning, what automatically
will turn to zero the corresponding canonical momenta. Naturally,
this receipt cannot be used if we want to account for the
fluctuations of the microscopic variables.

\section{Hamiltonian of the spin-isospin rotation} \label 3

Let us consider the dynamic solitons in the SU(2) field theories.
The initial properties of symmetry of these systems and the
corresponding integrals of motion play a key role in the
collective description of them. It is convenient to present the
Lagrangian reflecting these properties of symmetry in the form
\begin{equation}
L(\phi ,\dot{\phi})=1/2<g_{ik}(\phi )\dot{\phi}^{i}\dot{\phi}^{k}>-H(\phi
,0)\,,\qquad <A>\equiv \int d^{3}x\ A(x)\,.  \label{a}
\end{equation}
Here the field $\phi ^{i}(x,t)$ belongs to the space of the SU(2) group
parameters. This Lagrangian can describe the SU(2) sigma-model as well as the
corresponding Skyrme model depending on the concrete form of the kinetic
matrix $g_{ik}(\phi )$ and the potential term $H(\phi ,0)$. It is only
important to stress that the Lagrangian (\ref a) includes the mass term
providing for the existence of dynamic soliton states under sufficiently
moderate rotation. In the presence of this term the chiral invariance is
broken and the actual symmetry properties of the Lagrangian are reduced to
its invariance with respect to isotopic rotations $\phi(x)\rightarrow T\phi
(x)$ and to the invariance with respect to spatial rotations
$\phi (x)\rightarrow \phi (Tx)$. Here $T$ are the three-dimensional
orthogonal matrices. The conserved dynamic functionals of the isotopic and
ordinary spin associated with the invariance properties mentioned have the
form
\begin{equation}
\begin{split}
&\boldsymbol{I}=-i<\hat{\boldsymbol{I}}\phi
^{i}g_{ik}\dot{\phi}^{k}>,\qquad
\hat{I}_{ks}^{i}=-i\varepsilon _{iks}\ , \\
&\boldsymbol{J} =-i<\hat{\boldsymbol{l}}\phi
^{i}g_{ik}\dot{\phi}^{k}>,\qquad \hat{l}_{ks}^{i}=-i\delta
_{ks}(\boldsymbol{x}\times \partial /\partial
\boldsymbol{x})_{\,i}\,, \label{d}
\end {split}
\end{equation}
where $\hat{\boldsymbol{I}}$, $\hat{\boldsymbol{l}}$ are the
kinematic operators of the isotopic spin and the angular momentum.
The invariance of the Lagrangian with respect to the spatial
translations is not important in the problem considered and it
will not be dealt with further. In the general case a localized
soliton state is a dynamic system with a finite number of degrees
of freedom. The essential part of the problem is the determination
of the configurational space of this system. Simplifying the
consideration, let us deal with the sector of the theory with the
unity topological charge and arbitrary values of the angular
momentum and the isotopic spin. This sector is elementary with
respect to other ones. The localized states in it are not split
into inner fragments that may possess their own degrees of freedom
including the rotational ones. It may be expected that the
lifetime of such states due to the emission of mesons exceeds
considerably the times of inner motions associated with the
soliton rotation in the ordinary and isotopic spaces. In terms of
collective variables such evolution can be described with the
substitution
\begin{equation}
\phi (x)=T^{(i)}\varphi (T^{(s)-1}x)\,,  \label{b}
\end{equation}
that separates the cyclic variables of the system explicitly. In
this formula the group parameters $a_{\alpha}^{(i)}$,
$a_{\alpha}^{(s)}$  of the matrices $T^{(i)}$ and $T^{(s)}$ play
the part of the collective coordinates, describing the isotopic
and ordinary rotation of the soliton. The field $\varphi (x)$ is a
new generalized coordinate associated with the microscopic degrees
of freedom. As in this case the collective coordinates are the
cyclic ones, it is expedient again to obtain the equation for
dynamic solitons not making use of the general formula (\ref o).
Developing the canonical procedure on the ground of the
substitution (\ref b) and dealing with purely collective motions,
we omit the time derivative of the field $\varphi (x)$. According
to the prescription of the preceding section, such simplification
should be adjust with the gauge conditions. Below, in formulating
these conditions, we will identify the gauge function with the
coherent component of the field $\varphi (x)$. Thus,
differentiating the expression (\ref b) with respect to time
yields
\begin{equation}
\dot{\phi}(x)=-iT^{(i)}\left((\boldsymbol{\omega}^{(i)}\hat{\boldsymbol{I}}+
\boldsymbol{\omega}^{(s)}\hat{\boldsymbol{l}})\varphi(x)\right)_{|x\rightarrow
T^ {(s)-1}x}\,, \label{q}
\end{equation}
where $\boldsymbol{\omega }^{(i)}$,  $\boldsymbol{\omega }^{(s)}$  are the
left-invariant forms of angular velocities defined by the formulas
\begin{equation}
T^{(a)-1}\dot{T}^{(a)}=-i\hat{\boldsymbol{I}}\boldsymbol{\omega
}^{(a)}\ ,\qquad a=i,\ s\,.  \label{M}
\end{equation}
Inserting the formulas (\ref b), (\ref q) into the expression
(\ref a) can yield the effective Lagrangian
$L_{c}(a,\boldsymbol{\omega} ;\varphi (x))$. This Lagrangian does
not contain the velocities $\dot{\varphi}(x)$ and therefore it is
degenerate. In the generalized Hamiltonian formalism the field
$\varphi (x)$ is the variational parameter minimizing the
functional
\begin{equation}
H_{c}(\varphi )\equiv \underset{\nu =i,s}{\sum }\boldsymbol{\omega }^{(\nu )}%
\frac{\partial L_{c}}{\partial \boldsymbol{\omega }^{(\nu )}}-L_{c}  \label{r}
\end{equation}
with fixed values of other Hamiltonian variables. In order to pass
to these variables in the formula (\ref r), it is necessary to
employ the linear relations between the velocities
$\boldsymbol{\omega }^{(i)}$, $\boldsymbol {\omega }^{(s)}$ and
the conserved functionals $\boldsymbol{I}$, $\boldsymbol{J}$.
Using the formulas (\ref d), (\ref q), let us present these
relations in the matrix form
\begin{equation}
V^{\prime }=-\Lambda ^{\prime }\Omega\,,  \label{t}
\end{equation}
where the bi-vectors $\Omega $, $V^{\prime }$ are defined through the formulas
\begin{equation*}
V^{\prime }=\left(
\begin{array}{c}
\boldsymbol{I}^{r} \\
\boldsymbol{J}^{r}
\end{array}
\right) ,\qquad \Omega =\left(
\begin{array}{c}
\boldsymbol{\omega }^{(i)} \\
\boldsymbol{\omega }^{(s)}
\end{array}
\right).
\end{equation*}
Here $\boldsymbol{I}^{r}\equiv -\boldsymbol{I}T^{(i)}$, \ \
$\boldsymbol{J}^{r}\equiv -\boldsymbol{J}T^{(s)}$ are the dynamic
functionals having, according to (\ref s), the sense of the
generators of right shifts in the configurational space of the
group parameters \ $a_{\alpha }^{(i)}$, $a_{\alpha }^{(s)}$. The
kinetic matrix $\Lambda^{\prime }$  in (\ref t) has a block
structure. This matrix can be presented in the compact form as a
product of a column and a row
\begin{equation*}
\Lambda _{ik}^{\prime }(\varphi )=-\left\langle \left(
\begin{array}{c}
\hat{I}^{\,i} \\
\hat{l}^{\,i}
\end{array}
\right) \varphi ^{\,s}g_{st}(\varphi )(\hat{I}^{\,k},\,\hat{l}^{\,k})\varphi
^{\,t}\right\rangle\,.
\end{equation*}
From the viewpoint of the subsequent study of the axially
symmetric field configurations and close to them that are
characteristic for the states of the nucleon and delta, it is
expedient to use instead of the variables $\boldsymbol{I}^{r}$,
$\boldsymbol{J}^{r}$ their linear combinations
\begin{equation}
\boldsymbol{R}=(1/2)(\boldsymbol{I}^{r}-\boldsymbol{J}^{r}),\qquad
\boldsymbol{Q}=\boldsymbol{I}^{r}+ \boldsymbol{J}^{r}\,.
\label{rq}
\end{equation}
Introducing into consideration the bi-vector $\widetilde{V}\equiv
(\boldsymbol{R}, \boldsymbol{Q})$ and using the formula (\ref t)
yield the link between new variables $V$ and the velocities
$\Omega $
\begin{equation}
\Omega =-A\Lambda ^{-1}V\,,  \label{N}
\end{equation}
where the new kinetic matrix $\Lambda$ and the transformation matrix $A$ have
the form
\begin{equation*}
\Lambda (\varphi )=\widetilde{A}\Lambda ^{\prime }(\varphi )A=-\left(
\begin{array}{cc}
(DD) & (DK) \\
(KD) & (KK)
\end{array}
\right) ,\qquad A=\left(
\begin{array}{cc}
1/2 & 1 \\
-1/2 & 1
\end{array}
\right)\,.
\end{equation*}
The inner blocks of the matrix $\Lambda$ are defined through the formulas
\begin{equation}
\begin{array}{c}
(D^{i}D^{k})\equiv\,<\hat{D}^{i}\varphi ^{s}g_{st}(\varphi )\hat{D}%
^{k}\varphi ^{t}>, \\
(K^{i}D^{k})\equiv\,<\hat{K}^{i}\varphi ^{s}g_{st}(\varphi )\hat{D}%
^{k}\varphi ^{t}>,
\end{array}
\begin{array}{c}
(D^{i}K^{k})\equiv\,<\hat{D}^{i}\varphi ^{s}g_{st}(\varphi )\hat{K}%
^{k}\varphi ^{t}>, \\
(K^{i}K^{k})\equiv\,<\hat{K}^{i}\varphi ^{s}g_{st}(\varphi )\hat{K}%
^{k}\varphi ^{t}>, \label{G}
\end{array}
\end{equation}
where
\begin{equation*}
\boldsymbol{\hat{D}}=(1/2)(\boldsymbol{\hat{I}-\hat{l}}),\qquad
\boldsymbol{\hat{K}=\hat{I}+\hat{l}}
\end{equation*}
are the kinematic operators acting on the function $\varphi (x)$:
$\hat{D}^{i} \varphi ^{s}(x)\equiv
\hat{D}_{sb}^{i}\varphi^{b}(x)$. Thus in terms of the variables
$\boldsymbol{R}$, $\boldsymbol{Q}$ the functional $H_{c}(\varphi
)$ (\ref r) assumes the form
\begin{equation}
H_{c}(\varphi )=\frac{1}{2}\widetilde{V}\Lambda ^{-1}(\varphi )V+H(\varphi
,0)\,.  \label{J}
\end{equation}
In order to obtain the equation for coherent components of the
field $\varphi (x)$, we limit the admissible variations of this
field by the conditions
\begin{equation}
<\delta _{\varphi }\phi ^{q}g_{qk}(\phi )\frac{\partial }{\partial a^{(\nu )}%
}\phi ^{k}>=0\,,\qquad \nu =i,\,s\,.  \label{u}
\end{equation}
Here $\delta _{\varphi }\phi$ are the variations of the total
field $\phi (x)$ (\ref b) connected with the variations of the
function $\varphi(x)$. In accordance with the prescription of the
preceding section, the gauge function in (\ref u), associated with
the field $\varphi(x)$, is defined as its coherent component.
Under the signs of the trace and the integral the isotopic and
spin matrices $T$ in (\ref u) vanish. Besides, if one takes into
account the non-degeneracy of the matrices $\theta _{p\,\alpha }$,
defined with the formula
\begin{equation*}
T^{-1}\frac{\partial T}{\partial a_{\alpha }}=-i\hat{I}^{p}\theta
_{p\,\alpha}\,,
\end{equation*}
then the relations (\ref u) can be put in the form
\begin{equation*}
<\boldsymbol{\hat{D}}\varphi g\delta \varphi >=0,\qquad
<\boldsymbol{\hat{K}}\varphi g\delta \varphi >=0\,.
\end{equation*}
If one constructs on the directing vectors $\mathbf{\hat{D}}\varphi (x)$,
$\mathbf{\hat{K}}\varphi (x)$ the projection operator
\begin{equation}
\mathcal{P}\equiv -g(\hat{D},\hat{K})\varphi >\Lambda ^{-1}<\left(
\begin{array}{c}
\hat{D} \\
\hat{K}
\end{array}
\right) \varphi ,\qquad \mathcal{P}^{2}=\mathcal{P},  \label{z}
\end{equation}
then the gauge conditions (\ref u) can be rewritten as
\begin{equation}
\delta \varphi (1-\mathcal{P})=\delta \varphi .  \label{x}
\end{equation}
With the account of these conditions the coherent components of the field
$\varphi (x)$, realizing the extremum of the functional $H_{c}(\varphi )$,
will be determined from the equation
\begin{equation}
(1-\mathcal{P})\frac{\delta H_{c}(\varphi )}{\delta \varphi }=0\,.  \label{v}
\end{equation}
The solutions of this equation depend on the dynamic variables
$\boldsymbol{R}$, $\boldsymbol{Q}$ as on the parameters $\varphi
_{c}(x)\equiv \varphi _{c}(x; \boldsymbol{R},\boldsymbol{Q})$. The
functional $H_{c}(\varphi )$ dependent on these solutions
determines the collective Hamiltonian of the spin-isospin rotation
\begin{equation}
H_{c}(\boldsymbol{R},\boldsymbol{Q})\equiv H_{c}(\varphi _{c})\,.
\label{D}
\end{equation}
Let us make explicit the action of the operator $\mathcal{P}$ in
equation (\ref v). The functional $H_{c}(\varphi )$ as a function
of the dynamic variables $\boldsymbol{I}^{r}\boldsymbol{\equiv
-I}T^{(i)}$, $\boldsymbol{J}^{r}\boldsymbol{\equiv -J}T^{(s)}$
possesses an obvious property
\begin{equation}
H_{c}\left(\boldsymbol{I}^{r},\boldsymbol{J}^{r};\varphi
(x)\right)=H_{c}\left(\boldsymbol{I},\boldsymbol{J};T^{(i)}\varphi
(T^{(s)-1}x)\right)\,. \label{w}
\end{equation}
This relationship is valid for arbitrary $\varphi(x)$, not limited
by any additional conditions. Therefore, differentiating it with
respect to the parameters $a_{\alpha }^{(i)}$, $a_{\alpha
}^{(s)}$, one can find the identities relating the conventional
variational derivatives $\delta H_{c}/ \delta \varphi(x)$ with the
derivatives of the type $\partial H_{c}(\boldsymbol{I}
^{r},\boldsymbol{J}^{r};\varphi (x))/\partial \boldsymbol{I}^{r}$.
The latter derivatives can be regarded as complete on the
solutions of the equation (\ref v), i.e., taking into account the
dependence of the solutions $\varphi _{c}(x)$ on the parameters
$\boldsymbol{I}^{r}$, $\boldsymbol{J}^{r}$. It is possible,
because under such treatment the variational derivative of the
collective Hamiltonian with respect to $\varphi _{c}(x)$ in the
left-hand side of (\ref w) should be regarded with the account of
the coupling (\ref x), and therefore it falls out due to the
equations (\ref v). Thus we obtain from (\ref w) that
\begin{equation}
\int dx\boldsymbol{\hat{I}}\varphi (x)\frac{\delta H_{c}}{\delta
\varphi (x)}=i\dot{\boldsymbol{I}}^{r}, \qquad
 \int dx\boldsymbol{\hat{l}}\varphi (x)\frac{\delta
H_{c}}{\delta \varphi (x)}=i\dot{\boldsymbol{J}}^{r}. \label{A}
\end{equation}
Here $\dot{\boldsymbol{I}^r}\equiv \{\boldsymbol{I}^{r},H_{c}\}$,
$\dot{\boldsymbol{J}^r}\equiv \{\boldsymbol{J}^{r},H_{c}\}$ are
the rates of change of the variables $\boldsymbol{I}^{r}$,
$\boldsymbol{J}^{r}$ under the action of the collective
Hamiltonian $H_{c}(\varphi_c )$. The Poisson brackets of the
functionals $\boldsymbol{I}^{r}$, $\boldsymbol{J}^{r}$ are defined
with the expressions (\ref y). With the account of the formulas
(\ref z), (\ref A) the equation (\ref v) will ultimately assume
the form
\begin{equation}
\frac{\delta H_{c}(\varphi )}{\delta \varphi^n (x)}=-ig_{n\,l}(\varphi )(\hat{D},%
\hat{K})_{ls}^i \varphi^s (x)\Lambda _{ik}^{-1}\dot{V}^{k}\,.
\label{B}
\end{equation}
This equation determines the shape of the rotating soliton. The
right-hand part of this equation allows for the self-consistent
collective forces reflecting the availability of non-compensated
dynamic stresses in the system.

\section{Self-similar solutions of evolution equations} \label 4

We have defined the dynamic soliton states as the solutions of the
equation (\ref B). How are these solutions related to the
solutions of the exact evolution equations? If the ansatz (\ref b)
is substituted into the Lagrange equation of motion
\begin{equation*}
\frac{d}{dt}g(\phi (x))\dot{\phi}(x)=\frac{\delta }{\delta \phi (x)}\left(
\frac{1}{2}<\dot{\phi}g\dot{\phi}>-H(\phi ,0)\right)
\end{equation*}
with the assumption that $\varphi (x)$ in (\ref b) does not depend
on time and the conservation laws (\ref d) are taken into account,
then we obtain exactly the equation (\ref B) for the field
$\varphi (x)$. But the solutions of this equation depend on the
functionals $\boldsymbol{R}, \boldsymbol{Q}$, which are not the
integrals of motion, generally speaking. We will assume that in
the systems under consideration a spontaneous breaking of symmetry
takes place. Under this breaking the spin and isospin variables
entangle and the solutions of the equation (\ref B) remain
invariant with respect to arbitrary three-dimensional rotations:
\begin{equation}
\varphi _{c}(x;\boldsymbol{R},\boldsymbol{Q})=T^{-1}\varphi _{c}
(Tx;T\boldsymbol{R},T\boldsymbol{Q})\,.  \label{C}
\end{equation}
In favor of this choice there speaks the fact that in the static limit such
solutions are transformed into hedgehog configurations realizing the minimum
energy of one-baryon states in the Skyrme model. On the solutions (\ref C) the
collective Hamiltonian (\ref D) becomes the function of three invariant
functionals
\begin{equation}
H_{c}=H_{c}(\boldsymbol{R}^{2},\boldsymbol{Q}^{2}\boldsymbol{,RQ})\,.
\label{F}
\end{equation}
In agreement with the formulas (\ref E), this means that the
vector $\boldsymbol{R}$ is precessing uniformly around the vector
$\boldsymbol{Q}$ conserved in time according to the equation
\begin{equation*}
\dot{\boldsymbol{R}}\equiv \{\boldsymbol{R},H_{c}\}=\alpha
\boldsymbol{Q}\times \boldsymbol{R},\qquad \alpha \equiv
2\frac{\partial H_{c}}{\partial
\boldsymbol{Q}^{2}}-\frac{1}{2}\frac{\partial H_{c}}{\partial
\boldsymbol{R}^{2}}\,.
\end{equation*}
With arbitrary $\boldsymbol{R}, \boldsymbol{Q}$ the solutions of
the equation (\ref B) do not possess the axial symmetry and depend
on time through the parameter $\boldsymbol{R}$. Consequently, the
corresponding ansatz (\ref b) is not a solution of the equation of
motion, but it describes the true resonant states. In case of
axial symmetry (the states with $\boldsymbol{Q}=0$ have physical
sense, see below) the vector $\boldsymbol{R}$ is conserved in
time, and therefore the ansatz (\ref b) constructed on the
solutions of the equation (\ref B) becomes the exact solution of
the motion equations. The self-consistent forces vanish in this
case, and the equation (\ref B) itself is reduced to the equation
for the stationary states of the Hamiltonian
$H_{c}(\boldsymbol{R}^{2},0, 0)$ (\ref F).

\section{Quantizing the spin-isospin rotation} \label 5

In the quantum case the functionals $\boldsymbol{R}^{2}$,
$\boldsymbol{Q}^{2}$, $\boldsymbol{RQ}$ should be substituted with
the corresponding operators. These operators are equivalent to the
set of the operators $\boldsymbol{I}^{2}$, $\boldsymbol{J}^{2}$,
$\boldsymbol{I}^{r}\boldsymbol{J}^{r}$. The latter means that in
contrast to the spin  $\boldsymbol{J}$ and the isotopic spin
$\boldsymbol{I}$ , the generators of the right shifts
$\boldsymbol{J}^{r}$, $\boldsymbol{I}^{r}$ are not the integrals
of motion of the Hamiltonian (\ref F). Only their sum
$\boldsymbol{Q=I}^ {r}+\boldsymbol{J}^{r}$ is conserved.
Consequently, the eigenvalues $Q(Q+1)$, $q$, $I(I+1)$, $n$,
$J(J+1)$, $m$ of the complete set of the mutually permutable
operators $\boldsymbol{Q}^{2}$, $Q_{3}$, $\boldsymbol{I}^{2}$,
$I_{3}$, $\boldsymbol{J}^{2}$, $J_{3}$ can be used as the quantum
numbers enumerating the eigenstates of the Hamiltonian $H_c$.
Thus, the wave functions of the collective motion corresponding to
the noted quantum numbers can be presented in the form
\begin{equation}
|Q,q;I,n;J.m>\equiv \underset{m^{r}+n^{r}=q}{\sum }<I,n^{r};J,m^{r}|Q,q>\chi
_{n,n^{r}}^{I}(a^{(i)})\chi _{m,m^{r}}^{J}(a^{(s)})\,.  \label{in}
\end{equation}
Here $<I,n^{r};J,m^{r}|Q,q>$ are the angular coefficients, $\chi
_{n,n^{r}}^ {I}(a^{(i)})$ and $I(I+1), n, n^{r}$ are the
eigenstates and the respective eigenvalues of the isotopic
operators $\boldsymbol{I}^{2}, I_{3}, I_{3}^{r}$ . The same sense
is also attributed to the wave function of the spin motion $\chi
_{m,m^{r}}^{J}(a^{(s)})$. The invariant operators in (\ref F)
possess the following eigenvalues
\begin{eqnarray*}
\boldsymbol{RQ} &=&\frac{1}{2}(I(I+1)-J(J+1)), \\
\boldsymbol{Q}^{2} &=&Q(Q+1),\qquad |I-J|\leq Q\leq I+J, \\
\boldsymbol{R}^{2} &=&\frac{1}{2}(I(I+1)+J(J+1)-\frac
{1}{2}Q(Q+1)).
\end{eqnarray*}
In the sector with the baryon charge $B=1$ the quantization scheme
with half-integer spin and isospin values should be used \cite{W}.
In this case the eigenvalues of the operator $\boldsymbol{R}^{2}$
do not vanish, in contrast to the operators $\boldsymbol{Q}^{2}$,
$\boldsymbol{RQ}$. So the wave function of the nucleon with
$I=J=1/2$, $Q=0$ will correspond to the ground state. According to
Eq.(\ref {in}) this wave function has the form
\begin{equation}
|n,m>=\frac{1}{\sqrt{2}}\left( \chi _{n\uparrow }(a^{(i)})\chi _{m\downarrow
}(a^{(s)})-\chi _{n\downarrow }(a^{(i)})\chi _{m\uparrow }(a^{(s)})\right).
\label{O}
\end{equation}
Here $n,m=\pm 1/2$ are the projections of the isotopic and ordinary spins on
the quantization axis. The explicit expressions for the components of the
vector $\chi _{ks}^{1/2}(a)\equiv \chi _{ks}(a)$ in (\ref O) have the form
\cite{ANW}
\begin{equation}
\begin{split}
&\chi _{\uparrow \uparrow }(a) =\frac{1}{\pi }(a_{1}+ia_{2}),\qquad \chi
_{\uparrow \downarrow }(a)=-\frac{i}{\pi }(a_{0}-ia_{3}), \\
&\chi _{\downarrow \uparrow }(a) =\frac{i}{\pi }(a_{0}+ia_{3}),\qquad \chi
_{\downarrow \downarrow }(a)=-\frac{1}{\pi }(a_{1}-ia_{2}).  \label{Q}
\end{split}
\end{equation}
The wave functions of the nucleon (\ref O) differ from the
respective wave function of the paper \cite{ANW}. With the
approach considered the spin and isospin rotations are
independent, therefore they require a double set of group
parameters for their description. The classical states with $Q=0$
will be described by the axially symmetric configurations and
besides the nucleon they will include the delta-isobar states.
Depending on the meson mass which limits the possible values of
the angular momentum $\boldsymbol{R}^{2}$, other resonances with
$I=J$, $Q=0$ can also manifest themselves. In order to determine
the spectrum of these states it is necessary to treat the equation
(\ref B) in the axially symmetrical case with
$\boldsymbol{R}\neq0$, $\boldsymbol{Q}=0$ .

\section{O(2)-invariant configurations} \label 6

In the general case with arbitrary values of the dynamical
variables $\boldsymbol{R}$, $\boldsymbol{Q}$ the kinetic matrix
$\Lambda$, dependent on the solutions of the equation (\ref B), is
nondegenerate. If these variables become collinear or one of them
vanishes, the system acquires the axially symmetric configuration
and the projection of the tangent vector
$\boldsymbol{\hat{K}}\varphi _{c}(x)$ on the axis of symmetry
$\boldsymbol{k}$ vanishes: $\boldsymbol{k\hat{K}}\varphi
_{c}(x)=0$. It means that the matrix of the moments of inertia
$\Lambda$ becomes degenerate on the O(2)-invariant configurations.
The bi-vector $(0, \boldsymbol{k})$ is its zero mode. Under these
conditions, the perturbation theory and the passage to the limit
procedure become ambiguous. They depend on the passage path and
can contain nonanalytic terms. In order to find the physically
acceptable branches of the solution and to determine the order of
the passage to the limit, let us consider the properties of the
matrix $\Lambda$ on the O(2)-invariant configurations in more
detail. Obviously, the solutions of the equation (\ref B) are even
functions of the vectors $\boldsymbol{R}$, $\boldsymbol{Q}$. That
is, in case of axial symmetry these solutions are even functions
of the unit vector $\boldsymbol{k}$ directed along the axis of
symmetry. This permits to present the elements of the block matrix
$\Lambda$ (\ref G) as follows
\begin{equation*}
(D^{i}D^{k}) =-(a\nu _{ik}+bk_{i}k_{k}), \quad (D^{i}K^{k})
=(K^{i}D^{k})=-c\nu _{ik}, \quad (K^{i}K^{k}) =-d\nu _{ik},
\end{equation*}
where $a, b, c, d$ are the scalar coefficients, $\nu _{ik}\equiv
\delta _{ik}- k_{i}k_{k}$. It is possible to determine the roots
of the characteristic equation $|\Lambda -\lambda |=0$:
\begin{eqnarray*}
\lambda _{1} &=&0,\qquad \lambda _{2}=b, \\
\lambda _{3} &=&\lambda _{4}=\frac{1}{2}\left( d+a+\sqrt{(d-a)^{2}+4c^{2}}%
\right) , \\
\lambda _{5} &=&\lambda _{6}=\frac{1}{2}\left( d+a-\sqrt{(d-a)^{2}+4c^{2}}%
\right),
\end{eqnarray*}
and to present the corresponding orthonormalized eigenfunctions of the matrix
$\Lambda$ in the form
\begin{eqnarray*}
\psi _{1} &=&\left(
\begin{array}{c}
0 \\
1
\end{array}
\right) \boldsymbol{k},\qquad \quad \quad \quad\,\, \psi
_{2}=\left(
\begin{array}{c}
1 \\
0
\end{array}
\right) \boldsymbol{k,} \\
\psi _{3} &=&\alpha _{3}\left(
\begin{array}{c}
c \\
\lambda _{3}-a
\end{array}
\right) \boldsymbol{p},\qquad \psi _{4}=\alpha _{3}\left(
\begin{array}{c}
c \\
\lambda _{3}-a
\end{array}
\right) \boldsymbol{p}^{\prime }, \\
\psi _{5} &=&\alpha _{5}\left(
\begin{array}{c}
c \\
\lambda _{5}-a
\end{array}
\right) \boldsymbol{p},\qquad \psi _{6}=\alpha _{5}\left(
\begin{array}{c}
c \\
\lambda _{5}-a
\end{array}
\right) \boldsymbol{p}^{\prime }.
\end{eqnarray*}
Here $\alpha_3, \alpha_5$ are the normalizing constants,
$\boldsymbol{p}$, $\boldsymbol{p}^{\prime }$ are the unit
orthonormalized vectors in the plane orthogonal to the axis of
symmetry, viz. $\boldsymbol{p} \boldsymbol{p}^{\prime }=
\boldsymbol{p} \boldsymbol{k}=\boldsymbol{p}^{\prime }
\boldsymbol{k}=0$. Assuming the axis of symmetry  $\boldsymbol{k}$
to be directed along the vector $\boldsymbol{R}$, let us take into
account the infinitesimal vector $\boldsymbol{Q}_{\bot }$ for
regularizing the eigenvalue $\lambda_1$. Then using the
eigenfunctions $\psi_r$ we can establish the asymptotic behavior
of the matrix $\Lambda^{-1}$ in the vicinity of the axially
symmetric configurations
\begin{equation}
\begin{split}
\Lambda ^{-1}&=\overset{6}{\underset{r=1}{\sum }}\lambda _{r}^{-1}\psi _{r}%
\tilde{\psi}_{r}= \\
&=\lambda _{1}^{-1}(\boldsymbol{Q}_{\bot })\left(
\begin{array}{cc}
0 & 0 \\
0 & 1
\end{array}
\right) \hat{k}\hat{k}+b^{-1}\left(
\begin{array}{cc}
1 & 0 \\
0 & 0
\end{array}
\right) \hat{k}\hat{k}+(\lambda _{3}^{-1}A_{3}+\lambda _{5}^{-1}A_{5})\nu\,.
\label{I}
\end{split}
\end{equation}
Here $\lambda _{1}(\boldsymbol{Q}_{\bot })$ is the regularized
eigenvalue $\lambda_1$, tending to zero with $\boldsymbol{Q}_{\bot
}\rightarrow 0$; $A_3$, $A_5$ are the degenerate two-dimensional
matrices, whose form is inessential in this case. On inserting the
expression (\ref I) into the kinetic term $\widetilde{V}\Lambda
^{-1}V$ , the singular term survives only at $Q_{||}\equiv
\boldsymbol{kQ}\neq 0$. This means that the spectrum of classical
resonances corresponding to $Q_{||}\neq 0$  possesses a
singularity in the asymptotic range of small $\boldsymbol{Q}_{\bot
}^{2}$. Hence it also follows that for the description of the
states corresponding to the nucleon and delta ($\boldsymbol{Q}=0$)
one should first perform the passage to the limit
$\boldsymbol{Q}_{||}\rightarrow 0$, that will remove the singular
term. Only then one should put $\boldsymbol{Q}_{\bot }=0$. As a
result, the functional $H_{c}(\varphi)$ (\ref J) will reduce to
\begin{equation}
H_{c}(\boldsymbol{R},\varphi )=\frac{1}{2}\frac{\boldsymbol{R}^{2}}{b(\varphi )}%
+H(\varphi ,0)\,,  \label{K}
\end{equation}
where the functional of the moment of inertia $b(\varphi)$ is defined by the
formula
\begin{equation*}
b(\varphi )\equiv
-k_{i}k_{r}(D^{i}D^{r})=-<\boldsymbol{k\hat{I}}\varphi
^{s}g_{st}(\varphi )\boldsymbol{k\hat{I}}\varphi ^{t}>.
\end{equation*}
The self-consistent forces in the equation (\ref B) vanish with
$\boldsymbol{Q} \rightarrow 0$ for any order of the passage to the
limit. So the equation for the axially symmetric configurations
$\boldsymbol{R}\neq 0$, $\boldsymbol{Q}=0$ will assume the form of
the conventional equation for the stationary states of the
Hamiltonian (\ref K)
\begin{equation}
\frac{\delta }{\delta \varphi (x)}\left( \frac{1}{2}\frac{\boldsymbol{R}^{2}}{%
b(\varphi )}+H(\varphi ,0)\right) =0\,.  \label{L}
\end{equation}
One should seek the solutions of this equation among the class of functions
with the topological charge $B=1$. If one uses the exponential parametrization
of SU(2) group: $U(x)=\exp i\boldsymbol{\varphi }(x)\boldsymbol{\tau }$, then
the corresponding boundary conditions will have the form
\begin{equation*}
|\boldsymbol{\varphi }(\infty )|=0,\qquad |\boldsymbol{\varphi }(0)|=\pi.
\end{equation*}
That is, these conditions coincide with those for the profile
function in case of the static hedgehog configuration.

As the functional $\boldsymbol{R}$ at $\boldsymbol{Q}=0$ is the
integral of motion, then the configuration
$\varphi(x,\boldsymbol{R})$ determined by the equation (\ref L) is
conserved in time. It means that the corresponding total field
(\ref b) is the self-similar solution of the equation of motion.
In this solution only the group parameters $a_{\alpha}^{(i)}$,
$a_{\alpha}^{(s)}$ or, what is the same, the matrices $T^{(i)}$,
$T^{(s)}$ depend on time. This dependence can be found directly
using the integrals of motion (\ref d). From the definition of the
velocity forms (\ref M) it follows that
\begin{eqnarray*}
\dot{T}^{(i)} &=&-iT^{(i)}\,\boldsymbol{\hat{I}\omega}^{(i)}, \\
\dot{T}^{(s)}&=&-iT^{(s)}\,\boldsymbol{\hat{I}\omega}^{(s)}.
\end{eqnarray*}
Expressing with the help of the formula (\ref N) the velocities
$\boldsymbol {\omega }^{(i)}$, \ $\boldsymbol{\omega }^{(s)}$
through the functionals $\boldsymbol{R}$, $\boldsymbol{Q}$, let us
write these equations in the matrix form
\begin{equation*}
\left(
\begin{array}{c}
\dot{T}^{(i)} \\
\dot{T}^{(s)}
\end{array}
\right) =i\left(
\begin{array}{cc}
T^{(i)} & 0 \\
0 & T^{(s)}
\end{array}
\right) \,\boldsymbol{\hat{I}}\,A\,\Lambda ^{-1}V.
\end{equation*}
Inserting here the regularized expression for $\Lambda^{-1}$ (\ref
I), and performing first the passage to the limit
$\boldsymbol{Q}_{||}\rightarrow 0$ and only then putting
$\boldsymbol{Q}_{\bot}=0$ yield the equation for the rotation
matrices in case of axial symmetry
\begin{equation*}
\left(
\begin{array}{c}
\dot{T}^{(i)} \\
\dot{T}^{(s)}
\end{array}
\right) =\frac{i}{2b}\left(
\begin{array}{c}
T^{(i)} \\
-T^{(s)}
\end{array}
\right)\,\boldsymbol{\hat{I}R}\,.
\end{equation*}
Hence we find that
\begin{equation*}
T^{(i)}(t)=T^{(s)}(-t)=\exp
\left(\frac{i}{2\,b}\,\boldsymbol{\hat{I}R}\, t\right).
\end{equation*}
That is, the isotopic and spin subsystems of the classical
localized state are rotating uniformly in opposite directions with
the angular frequency $\omega = |\boldsymbol{R}\,|/2b$.

Inserting the solutions of the equation (\ref L) into the
functional (\ref K) can determine the classical Hamiltonian of the
system on the surface $\boldsymbol{Q}=0$ for arbitrary
$\boldsymbol{R}^2$ values. At the points $I=J=1/2$ and $I=J=3/2$
this Hamiltonian determines the exact nucleon and delta masses in
the model considered. If the solutions of the equation (\ref L)
are many-valued, then in addition to the nucleon and delta this
equation can describe other resonances from the series $P_{11}$,
$P_{33}$. Besides, depending on the meson mass, the spectrum of
the solutions of this equation can also contain resonances with
$I=J>3/2$, $Q=0$. The complete picture of solutions of the
equation (\ref L) can be given only by means of its numerical
analysis. In the $\boldsymbol{R}^2$ approximation the Hamiltonian
(\ref K) reduces to the Hamiltonian of the adiabatic approach
\cite{ANW} for the Skyrme model. The equation (\ref L) makes a
strict account of the rotational deformation, and, by doing so,
lowers the nucleon and delta masses in comparison with the results
of the paper \cite{ANW}. In particular, one can show that the term
of the Hamiltonian (\ref K) of the order $\boldsymbol{R}^4$ is
negative. In the sigma-model there are no static soliton solutions
and the nucleon mass has a purely quantum origin being due to the
spin-isospin rotation of the soliton.

Let us employ the conventional scale analysis and prove that the
presence in the formula (\ref K) of the kinetic term
\begin{equation*}
H_{T}(\varphi )\equiv
\frac{1}{2}\frac{\boldsymbol{R}^{2}}{b(\varphi )}
\end{equation*}
impedes the collapse of the rotating soliton state in the SU(2)
sigma-model. In this model the potential $H(\varphi,0)$ has the
following form
\begin{equation*}
H(\varphi ,0)=H_{0}(\varphi )+H_{m}(\varphi )\,,
\end{equation*}
where $H_{0}(\varphi)$ is the conventional second-order term in
derivatives, $H_{m}(\varphi)$ is the mass term not containing the
spatial derivatives. As in the sigma-model the matrix $g(\varphi)$
also does not contain the spatial derivatives, then as a result of
the substitution $\varphi(x)\rightarrow \varphi(\lambda x)$ the
Hamiltonian (\ref K) is reduced to the form
\begin{equation*}
H_{c}(\boldsymbol{R},\varphi )\rightarrow H_{c\lambda }=\lambda
^{3}H_{T}+\lambda ^{-1}H_{0}+\lambda ^{-3}H_{m}\,.
\end{equation*}
It follows from this identity considered for the solutions of the equation
(\ref L) that
\begin{eqnarray*}
\frac{\partial H_{c\lambda }}{\partial \lambda }_{|\lambda =1} &=&0\quad
\rightarrow \quad H_{T}=H_{m}+\frac{1}{3}H_{0}\,, \\
\frac{\partial ^{2}H_{c\lambda }}{\partial \lambda ^{2}}_{|\lambda =1}
&>&0\quad \rightarrow \quad 2H_{T}+H_{m}>0\,.
\end{eqnarray*}
That is, the soliton energy $E=H_{T}+H_{0}+H_{m}$ can be finite
only in the presence of the kinetic term $H_{T}>0$. Thus, if a
soliton can rotate keeping the angular momentum $\boldsymbol{R}$,
it will be an object stable with respect to the collapse. In order
to estimate the possibility of such a rotation, let us determine
the asymptotics of the field $\varphi(x)$ at the spatial infinity.
On linearizing the equation (\ref L) with respect to $\varphi(x)$
it assumes the form
\begin{equation*}
\left( -\Delta +m_{\pi }^{2}-\left(
\frac{\boldsymbol{R}}{b}\right) ^{2}\nu \right) \varphi (x)=0\,.
\end{equation*}
Here $m_{\pi }$ is the meson mass, $\nu _{ik}\equiv \delta
_{ik}-k_{i}k_{k}$. This equation is valid for the sigma-model as
well as for the Skyrme model, because at small $\varphi(x)$ the
contribution of the Skyrme term falls out. Thus the field
component in the rotation plane $\nu\boldsymbol {\varphi}(x)$ will
have the damping asymptotics only for $\boldsymbol{R}^{2}<b^{2}
m_{\pi }^{2}$. The moment of inertia $b$ also depends on
$\boldsymbol{R}^2$, and therefore one can determine the exact
boundary of admissible values of the moment $\boldsymbol{R}$, at
which the soliton remains stable with respect to the
disintegration only on the ground of the direct numerical analysis
of the equation (\ref L). Thus in the sigma-model, in contrast to
the static case, there are conditions for the existence of the
dynamic soliton that is stable against the collapse and whose
shape is determined from the equation (\ref L).

\section{Conclusion} \label 7

The paper formulates the exact equations for the SU(2)-dynamic
solitons performing the rotations in the ordinary as well as
isotopic space. Generally, the shape of these solitons is
determined with the account of the self-consistent collective
forces reflecting the presence of non-compensated dynamic stresses
in the system. On assuming the spontaneous breaking of the
symmetry, the eigenfunctions of the Hamiltonian of the
spin-isospin rotation in the one-baryon sector are established.
Within the quantization scheme with a half-integer spin, the
axially symmetric field configurations correspond to the ground
state. The generalized matrix of the moments of inertia is
degenerate on such configurations. With the account of this
circumstance, the order of the passage to the limit near
singularities is established, and the equation for the
O(2)-invariant solitons is found. It is shown that, in contrast to
the general case, the O(2)-invariant configurations are equivalent
to the exact self-similar solutions of the equation of motion. The
temporal dependence of the respective dynamic states is
established. The quantum analogues of similar axially symmetric
solutions correspond to the states of the nucleon and delta. This
paper gives the expression for the state vector of the nucleon
reflecting the independent evolution of the soliton in the spin
and isospin spaces. The account of rotations is shown to create
the condition for the existence of the exact extended solutions of
the equations of motion in the SU(2) sigma-model. The treatment
performed is based on using the variational approach to the method
of collective variables. This paper gives the strict justification
of the equations of collective dynamics. 

\vspace{12pt}

\section*{Appendix}
\renewcommand{\theequation}{A.\arabic{equation}}
\setcounter{equation}{0}

The canonical description of the collective motion of the system
can be made in a standard way. The elements of the configurational
space of the collective subsystem are the parameters
$a_{\alpha}^{(i)}$,  $a_{\alpha}^{(s)}$ of the groups SO(3)
associated with the isotopic and spin rotation of the soliton.
With the help of the formulas (\ref a), (\ref d), (\ref q) the
canonical momenta conjugated to these variables can be related to
the functionals $\boldsymbol{I}^r$,  $\boldsymbol{J}^r$
\begin{equation}
\pi _{\alpha }^{(i)}\equiv \frac{\partial L}{\partial
\dot{a}_{\alpha}^{(i)} }=-\boldsymbol{I}^{r}\frac{\partial
\boldsymbol{\omega }^{(i)}}{\partial \dot{a}
_{\alpha }^{(i)}}\,,\qquad \pi _{\alpha }^{(s)}\equiv \frac{\partial L}{%
\partial \dot{a}_{\alpha }^{(s)}}=-\boldsymbol{J}^{r}\frac{\partial \boldsymbol
{\omega }^{(s)}}{\partial \dot{a}_{\alpha }^{(s)}}\, \label{P}\,.
\end{equation}
Obviously, the transition to the canonical description is
identical for the spin and isospin subsystems. Therefore, the
canonical properties of the isotopic subsystem can also be
extended to the spin variables. If the elements of the SU(2) group
are parametrized by the components of the unit vector on the
three-sphere: $A\equiv a_{0}+i\boldsymbol{\tau a}$, $a_{\alpha }^
{2}\equiv a_{0}^{2}+a_{k}^{2}=1$, $k=1\,,2\,,3$, then the
orthogonal matrix $T$ and the velocity form $\boldsymbol{\omega}$,
can be written as follows
\begin{eqnarray*}
T_{ik}&=&(1/2)trA\tau _{i}A^{-1}\tau
_{k}=(1-2\boldsymbol{a}^{2})\delta
_{ik}+2a_{i}a_{k}-2\varepsilon _{iks}a_{s}a_{0}\,, \\
\omega_{k}&=&-itr\dot{A}A^{-1}\tau _{k}=2(a_{0}\dot{a}_{k}-\dot{a}%
_{0}a_{k}+\varepsilon _{ksr}\dot{a}_{s}a_{r})\,.
\end{eqnarray*}
In order to determine the classical Poisson brackets of the
functionals $\boldsymbol{I}$, $\boldsymbol{I}^r$, it is convenient
to take the spatial components of the unit vector $a_{\alpha}$:
$a_k$, $k=1\,,2\,,3$ as the independent collective coordinates.
Noting that $\dot{a}_{0}=-a_{0}^{-1}a_{k}\dot{a}_{k}$, we find
from (\ref P) the following expressions for the functionals
$\boldsymbol{I}$, $\boldsymbol{I}^r$
\begin{eqnarray*}
I_{k}^{r} &=&\frac{1}{2}(\varepsilon_{kqr}a_{q}^{(i)}\pi
_{r}^{(i)}-a_{0}^{(i)}\pi_{k}^{(i)})\,, \\
I_{k} &=&\frac{1}{2}(\varepsilon_{kqr}a_{q}^{(i)}\pi
_{r}^{(i)}+a_{0}^{(i)}\pi_{k}^{(i)})\,.
\end{eqnarray*}
Hence, using the canonical Poisson bracket $\{a_{i},\pi
_{k}\}=\delta _{ik}$, we find the Poisson brackets for the
functionals $\boldsymbol{I}$, $\boldsymbol{I}^r$
\begin{eqnarray}
\{I_{s},\ I_{k}\} &=&\varepsilon _{skn}I_{n}\ ,\qquad \{I_{s}^{r},\
I_{k}^{r}\}=\varepsilon _{skn}I_{n}^{r}\ ,\quad \{I_{s},\ I_{k}^{r}\}=0\ ,
\label{y} \\
\{I_{k},\ T_{sn}\} &=&\varepsilon _{kst}T_{tn}\ ,\qquad \{I_{k}^{r},\
T_{sn}\}=\varepsilon _{ikn}T_{si}\,.  \label{s}
\end{eqnarray}
The last two formulas show that the functionals $\boldsymbol{I}$,
$\boldsymbol{I}^r$ are the dynamic generators of the left and
right shifts in the configurational space of the isotopic
variables. Using similar formulas for the respective functionals
of the spin subsystem $\boldsymbol{J}$, $\boldsymbol{J}^{r}$, we
find the Poisson brackets for the functionals $\boldsymbol{Q}$,
$\boldsymbol{R}$ (\ref {rq})
\begin{equation*}
\{Q_{i},Q_{j}\}=\varepsilon _{ijk}Q_{k}\ ,\qquad \{R_{i},R_{j}\}=\frac{1}{4}%
\varepsilon _{ijk}Q_{k}\ ,\qquad \{Q_{i},R_{j}\}=\varepsilon _{ijk}R_{k}\,,
\end{equation*}
as well as the Poisson brackets of these functionals with their
invariant combinations $\boldsymbol{R}^{2}$, $\boldsymbol{Q}^{2}$,
$\boldsymbol{RQ}$
\begin{equation}
\begin{split}
&\{Q_{i},\boldsymbol{RQ}\} =0\ ,\qquad
\{Q_{i},\boldsymbol{Q}^{2}\}=0\ ,\qquad
\{Q_{i},\boldsymbol{R}^{2}\}=0\ , \\
&\{R_{i},\boldsymbol{RQ}\} =0\ ,\qquad
\{R_{i},\boldsymbol{Q}^{2}\}=2\varepsilon _{iks}Q_{k}R_{s}\
,\qquad \{R_{i},\boldsymbol{R}^{2}\}=\frac{1}{2}\varepsilon
_{iks}R_{k}Q_{s}\ .  \label{E}
\end{split}
\end{equation}
In case of quantum description all components of the unit vector
on the three-sphere are conveniently considered as the collective
coordinates. Covariant usage of all components of the unit
four-vector permits to put the eigenfunctions of the left and
right generators in their most symmetric form. Defining the
canonical momenta $\pi _{\alpha }^{\prime }$, $\alpha
=0,\,1,\,2,\,3$ corresponding to these components, we find from
(\ref P) the following relations
\begin{eqnarray*}
I_{k}^{r} &=&\frac{1}{2}(\pi _{0}^{\prime (i)}a_{k}^{(i)}-a_{0}^{(i)}\pi
_{k}^{\prime (i)}+\varepsilon _{kqr}a_{q}^{(i)}\pi _{r}^{\prime (i)})\,, \\
I_{k} &=&\frac{1}{2}(a_{0}^{(i)}\pi _{k}^{\prime (i)}-\pi _{0}^{\prime
(i)}a_{k}^{(i)}+\varepsilon _{kqr}a_{q}^{(i)}\pi _{r}^{\prime (i)})\,.
\end{eqnarray*}
With the account of these solutions the equation (\ref P) for $\pi_0$ is
satisfied identically. Considering the momenta $\pi _{\alpha}^{\prime}$ as
the operators on the three-sphere: $\pi _{\alpha }^{\prime }=-i\partial /
\partial a_{\alpha }$, one can establish the commutator relations for the
operators $\boldsymbol{I,\,\,I^r}$, corresponding to the Poisson
brackets (\ref y) and find the eigenfunctions of these operators
(\ref Q) \cite{ANW}.

\vspace{0,5 cm}

\end{document}